\documentclass[aps,prl,twocolumn]{revtex4-1}

\usepackage{amsmath}
\usepackage{graphicx}
\usepackage{color}

\newcommand{\Rey}{\mbox{Re}}
\newcommand{\Pran}{\mbox{Pr}}
\newcommand{\Nu}{\mbox{Nu}}

\newcommand{\Ra}{\mbox{Ra}}
\renewcommand{\vec}[1]{\boldsymbol{\mathbf{#1}}}
\newcommand{\uvec}[1]{\vec{\hat{#1}}}

\begin{document}


\title{Elliptical Instability and Multi-Roll Flow Modes of the Large-scale Circulation\\ in Confined Turbulent Rayleigh--B\'enard Convection}


\author{Lukas Zwirner}
\email[]{lukas.zwirner@ds.mpg.de}
\affiliation{Max Planck Institute for Dynamics and Self-Organization, Am Fassberg 17, 37077 G\"ottingen, Germany}
\author{Andreas Tilgner}
\affiliation{Institute for Geophysics, Georg-August University of G\"ottingen, Friedrich-Hund-Platz 1, 37077 G\"ottingen, Germany}
\author{Olga Shishkina}
\email[]{olga.shishkina@ds.mpg.de}
\homepage[]{\\http://www.lfpn.ds.mpg.de/shishkina/index.html}
\affiliation{Max Planck Institute for Dynamics and Self-Organization, Am Fassberg 17, 37077 G\"ottingen, Germany }


\date{\today}

\begin{abstract}
Turbulent Rayleigh--B\'enard convection in slender cylindrical cells exhibits rich dynamics of the large-scale circulation (LSC), with several rolls stacked on top of each other.
We propose that the elliptical instability is the mechanism which causes the twisting and breaking of the LSC into multiple rolls and that the volume-averaged heat and momentum transport, represented by the Nusselt number and Reynolds number, is generally weaker for larger number $n$ of the LSC rolls.
This is supported by direct numerical simulations for $\Ra=5\times10^5$, $\Pran=0.1$, $H=5D$ and $1\leq~n\leq4$.

\end{abstract}

\maketitle



In thermally driven flows, one of the most prominent features is the large-scale circulation (LSC) of a fluid, which contributes significantly to the heat and mass transport in the system.
The capability of the LSC to transport heat and mass is influenced by its shape and rich dynamics.
Rayleigh--B\'enard convection (RBC), where a fluid is confined between a heated plate (at temperature $T_+$) from below and a cooled plate (at temperature $T_-$) from above, is a paradigmatic system in thermal convection studies \cite{Bodenschatz2000, Ahlers2009, Lohse2010, Chilla2012};
it is characterized by the Rayleigh number, $\Ra\equiv \alpha g \Delta H^3/(\kappa\nu)$ (thermal driving), the Prandtl number, $\Pran\equiv\kappa/\nu$ (fluid property), and the geometry of the convection cell
\footnote{
Here, $\alpha$ is the isobaric thermal expansion coefficient,
$\nu$ the kinematic viscosity,
$\kappa$ the thermal diffusivity,
$g$ the acceleration due to gravity,
$\Delta\equiv T_+-T_-$ and
$T_+$ and $T_-$  are the temperatures of heated and cooled plates.
}.

Although the LSC in RBC has been known for a long time, recent investigations aim to provide a deeper understanding of its versatile dynamics, e.g. reversals, precession, sloshing and twisting~\cite{Funfschilling2004, Funfschilling2008, Xi2009, Zwirner2020}.
One particular factor that influences the LSC is the geometry of the convection cell.
Several studies focused on how lateral confinement in one direction influences the heat transport and flow structures \cite{Wagner2013, Huang2013, Chong2016, Chong2015, Chong2018}, though only a few studies focused on lateral confinement in two directions, e.g. slender cylindrical cells of small diameter-to-height aspect ratio $\Gamma=D/H$.
Not only a single-roll mode (SRM) of the LSC, but also a double-roll mode (DRM) --- composed of two rolls on top of each other --- was found for cylindrical cells with $\Gamma=1$, $1/2$, $1/3$ and $1/5$ \cite{Xi2008, Weiss2013, Zwirner2018}.
Experimental studies with water ($\Pran\approx5$) found that the SRM is characterized by a slightly enhanced heat transport ($\approx 0.5\,\%$) compared to the DRM \cite{Xi2008, Weiss2013}.
It was also found that small-$\Gamma$ systems spend more time in the DRM than in the SRM.
Direct Numerical Simulations (DNS) \cite{Zwirner2018} for $\Ra=10^6$, $\Pran=0.1$ and $\Gamma=1/5$ showed that the heat transport of the DRM is only $80\,$\% compared to the SRM.

In 2D DNS \cite{Poel2011, Poel2012}, up to four vertically stacked rolls were found for $\Gamma=0.4$.
Less heat transport was observed in the case of more rolls and the comparison at different $\Pran$ revealed that the $\Gamma$-dependence is more pronounced at lower $\Pran$.
It remains unclear, however, whether in 3D there exist multi-roll flow modes (with three or more rolls on top of each other), what is their efficiency in heat transport and which mechanism creates these modes.

In this Letter, we explore multi-roll modes of the LSC in RBC and propose the elliptical instability as a plausible mechanism to trigger their formation~\cite{Kerswell2002}.
This inertial instability also plays a role, e.g., in the precession-driven motion of the Earth's core \cite{Lorenzani2003} and in the dynamics of vortex pairs \cite{Leweke1998, Leweke2016}.
In the study \cite{Bars2006} of the elliptical instability under an imposed radial temperature gradient, it was found that its growth rate decreases with increasing $\Ra$, but this effect is less pronounced at low $\Pran$.\\

\paragraph{Numerical method.}
We conduct DNS using the high-order finite-volume code {\sc goldfish} \cite{Kooij2018}, which solves the momentum and energy equations in Oberbeck-Bousinessq approximation, for an incompressible flow ($\nabla\cdot\vec{u}=0$):
\begin{eqnarray}
\partial_t \vec{u} + \vec{u}\cdot\nabla\vec{u} &=&-\nabla p+ \nu\nabla^2 \vec{u}+\alpha g\theta\uvec{z},\label{eq:NSu}\\
\partial_t \theta + \vec{u}\cdot\nabla \theta &=& \kappa\nabla^2 \theta,\label{eq:NST}
\end{eqnarray}
The DNS were conducted at $\Ra=5\times10^6$, $\Pran=0.1$ and $\Gamma=1/5$, using a mesh of $256\times128\times22$ nodes in $z$, $\varphi$ and $r$-directions, which is of sufficient resolution \cite{Shishkina2010, Zwirner2018}.
We consider the volume-averaged instantaneous heat transport $\Nu(t) \equiv (\langle u_z(t)\,\theta(t)\rangle - \kappa\langle\partial_z\theta(t)\rangle)/(\kappa\Delta/H)$ (Nusselt number) and the Reynolds number, $\Rey(t)\equiv H\langle\vec{u^2}\rangle/\nu$, which is based on the kinetic energy.
Here and in the following $\langle\cdot\rangle$ denotes volume average, $\overline{\ \cdot\ }$ time average and $\langle\cdot\rangle_S$ horizontal area average.\\




\begin{figure*}
\includegraphics{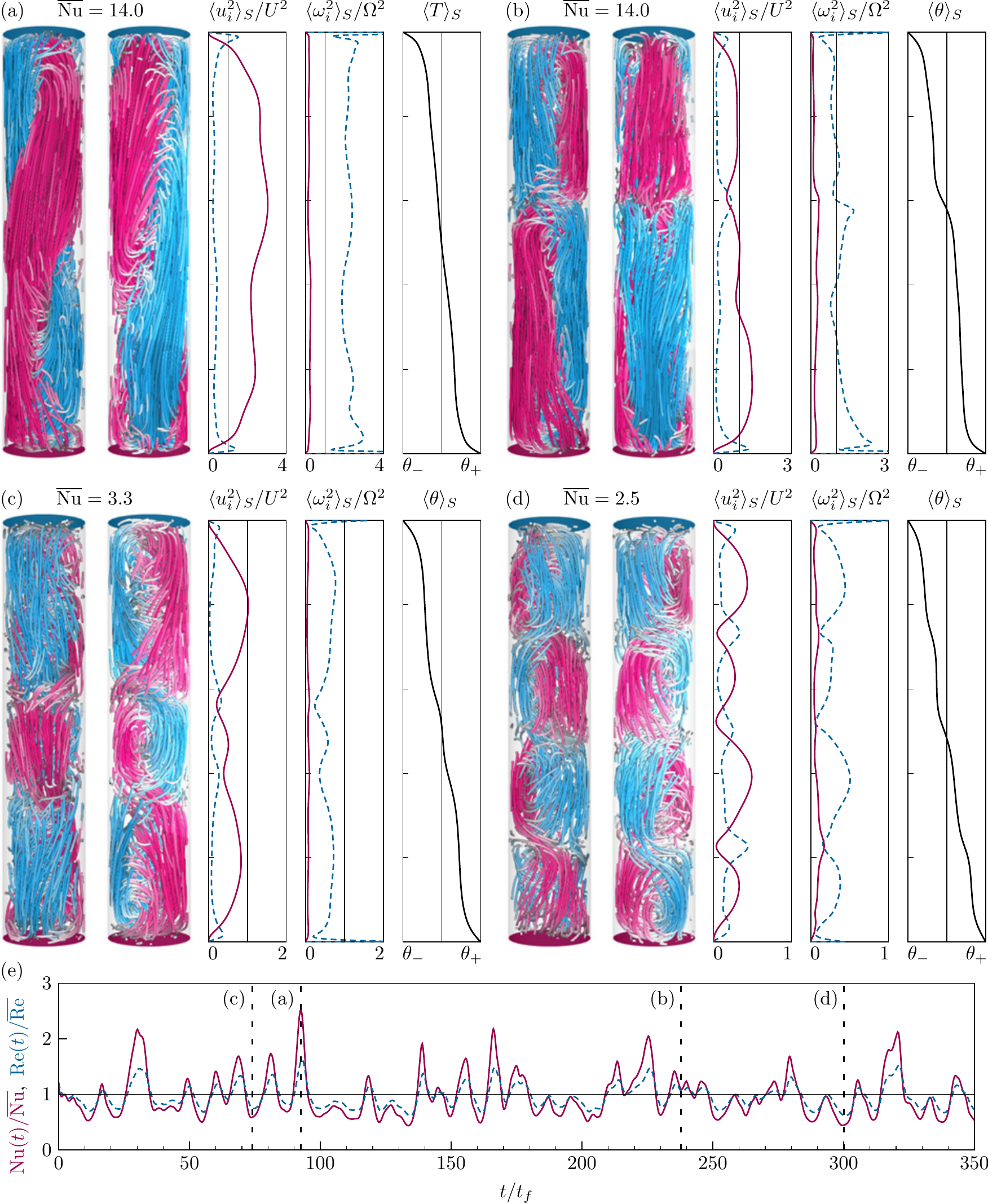}
\caption{Instantaneous flow fields, for a LSC composed of a different number $n$ of rolls: (a) $n=1$, (b) $n=2$, (c) $n=3$, (d) $n=4$.
Trajectories of passive tracer particles in two perpendicular perspectives, obtained with the ParaView \textit{"Particle Tracer"} filter (pink for upward and blue for downward flows), the normalized horizontally averaged profiles of the squared vertical (pink solid) and horizontal (dashed blue) components of the velocity $u_i$ and vorticity $\omega_i$ and the horizontally averaged profiles of the temperature $\theta$ are shown for each case.
(e) Temporal evolution of the normalized volume-averaged heat flux $\Nu(t)/\overline\Nu$.
The times of the snapshots (a)--(d) are marked by vertical dashed lines.
Parameters are: $\Ra=5\times10^6$, $\Pran=0.1$, $\Gamma=1/5$.}
\label{FIG1}
\end{figure*}

\paragraph{Properties of different flow modes.}
Inside the slender cylindrical cell of $\Gamma=1/5$, we observe flow modes consisting of up to $n=4$ distinct rolls, which are vertically stacked (Fig.~\ref{FIG1}a--d).
These $n$-roll flow modes endure for a few free-fall time units, $t_f\equiv H/\sqrt{\alpha g\Delta H}$, before they transition into another mode.
From time to time, the SRM is strongly twisted (Fig.~\ref{FIG1}a), before it breaks up into two distinct rolls (Fig.~\ref{FIG1}b).
Also, the rolls of the DRM may break up into more rolls or are only twisted for a certain time period.
Whether the rolls are twisted or break up can be distinguished by the shape of the profiles along the cylinder axis of different horizontally-averaged quantities, in particular, of the normalized  horizontal and vertical components of the squared velocity, $u^2_\text{h}(t, z)/U^2 = \langle u^2_r+u^2_\varphi\rangle_{S}/\overline{\langle\vec{u}\cdot\vec{u}\rangle}$ and  $u^2_\text{v}(t, z)/U^2 = \langle u^2_z\rangle_{S}/\overline{\langle\vec{u}\cdot\vec{u}\rangle}$, respectively, and of the temperature $\theta(t, z)$.
Note that the profiles are averaged over horizontal slices and depend on time~$t$ and the vertical coordinate~$z$.
In Fig.~\ref{FIG1}a--d the profiles are presented next to the corresponding snapshots of the flow modes.
Additionally, the enstrophy profiles, $\omega^2_i(t, z)$, are shown, which will be discussed below.
Note, that the profiles $u^2_\text{v}(z)$ and $u^2_\text{h}(z)$ of the DRM (Fig.~\ref{FIG1}b) have, respectively, a characteristic local minimum and maximum at the junction of the rolls ($z\approx 3/5\,H$) in contrast to the twisted SRM (Fig.~\ref{FIG1}a), where these extrema are absent.
Moreover, the temperature profile of the DRM shows a characteristic step-like behaviour at junction height;
there, the temperature gradient is locally increased, resembling a thermal boundary layer.
These shapes of the vertical profiles are characteristic and independent from the number of rolls.
Based on the analysis of these profiles, we developed an algorithm to extract the distinct rolls at any time step and performed conditional averaging on either each $n$-roll flow mode or each roll individually (see supplementary material for details).

Note that the rolls are not necessarily equally distributed within the cell.
Thus, two smaller rolls and one larger roll can form a three-roll mode.
This is similar to the findings for water~\cite{Xi2008}, where a DRM, consisting of a larger roll and a smaller one, was observed.
Although one might expect a five-roll mode for the aspect ratio $\Gamma=1/5$ as well, such a mode was not observed during the simulated time interval.
However, it cannot be excluded that a five-roll mode exists, as it is presumably a rare mode.
Note that Xi and Xia \cite{Xi2008} also did not observe a triple-roll mode in their cell of $\Gamma=1/3$.

Furthermore, we examine the enstrophy $\vec{\omega}^2$, which is the squared vorticity, $\vec{\omega} \equiv \nabla\times\vec{u}$, and splits similarly to the squared velocities, into the horizontal, $\omega^2_\text{h}=\omega_r^2+\omega_\varphi^2$ and vertical, $\omega^2_\text{v}=\omega_z^2$ contributions.
These are normalized with $\Omega^2=\overline{\langle\vec{\omega}\cdot\vec{\omega}\rangle}$.
The horizontal component of the enstrophy profile, $\omega_{h}^2(z)$, is strong within the region of a distinct roll and shows a local minimum at the juncture of two rolls and close to the cooled and heated plates (Fig.~\ref{FIG1}a--d).
On the other hand, the vertical component of the enstrophy is approximately one order of magnitude weaker (Tab.~\ref{TAB:roll-properties}).

As discussed above, the vertical profiles allow detection and systematic analysis of all $n$-roll flow modes. 
One of the primary quantities of interest in a thermally convective system is the global heat transport ($\Nu$).
Tab.~\ref{TAB:roll-properties} lists, among other quantities, the Nusselt number of each flow mode, and it shows that $\Nu$ decreases as the number of rolls increases.
This is a consistent extension of previous studies \cite{Xi2008, Weiss2013, Zwirner2018}, where only SRM and DRM were observed.
In contrast to high-$\Pran$ experiments \cite{Xi2008, Wei2013}, where the difference in the heat transport between the SRM and DRM was only $\approx0.5\,\%$, the decrease of $\Nu$ in the DRM is apparently much larger ($\approx 30\,\%$) at low $\Pran$.

Besides that, the heat transport also varies strongly in time (Fig.~\ref{FIG1}e), the standard-deviation of $\Nu$ is 2.6 and the distribution has a strong positive skewness (31.7), which means a long tail at high $\Nu$.
The Reynolds number, $\Rey$, varies less strongly with time (Fig.~\ref{FIG1}e).
The system is most likely to be in a DRM ($40.6\,\%$). 
Additionally, Tab.~\ref{TAB:roll-properties} gives the lifetimes, $\tau_n$ of each flow mode.
The mean lifetime of any flow mode is approximately $2\,t_f$.\\

\begin{table}
	\caption{Lifetimes $\tau_n$, probabilities $P_n$, mean heat transport $\Nu_n$, mean Reynolds number $\Rey_n$, horizontal $\overline{\langle\omega_\text{h}^2\rangle}$ and vertical enstrophy $\overline{\langle\omega_\text{v}^2\rangle}$, of the $n$-roll flow modes, for $\Ra=5\times10^6$, $\Pran=0.1$, $\Gamma=1/5$.}
	\label{TAB:roll-properties}
	\begin{ruledtabular}
		
		\begin{tabular}{ccccccc}
			$n$						& $\tau_n/t_f$			& $P_n/$\%	& $\Nu_n$		& $\Rey_n$	& $\overline{\langle\omega_\text{h}^2\rangle}t_f^2$& $\overline{\langle\omega_\text{v}^2\rangle}t_f^2$\\[2pt] \hline
			1						& $2.4\pm0.4$			& $30.1$	& $7.8\pm0.3$	& $990\pm30$		& $55\pm2$		& $7.0\pm0.3$		\\
			2						& $1.5\pm0.2$			& $40.6$	& $5.2\pm0.3$	& $820\pm20$		& $36\pm2$		& $5.9\pm0.3$		\\
			3						& $1.3\pm0.2$			& $24.0$	& $3.8\pm0.3$	& $720\pm20$		& $27\pm2$		& $5.7\pm0.3$		\\
			4						& $1.3\pm0.4$			& $5.3$ 	& $3.1\pm0.3$	& $640\pm30$		& $21\pm2$		& $4.8\pm0.3$		\\[2pt] \hline
			avg						& $1.6\pm0.1$			& ---		& $5.5\pm0.3$	& $830\pm20$		& $39\pm2$		& $6.2\pm0.3$		
		\end{tabular}
	\end{ruledtabular}
\end{table}

\begin{figure}
\includegraphics{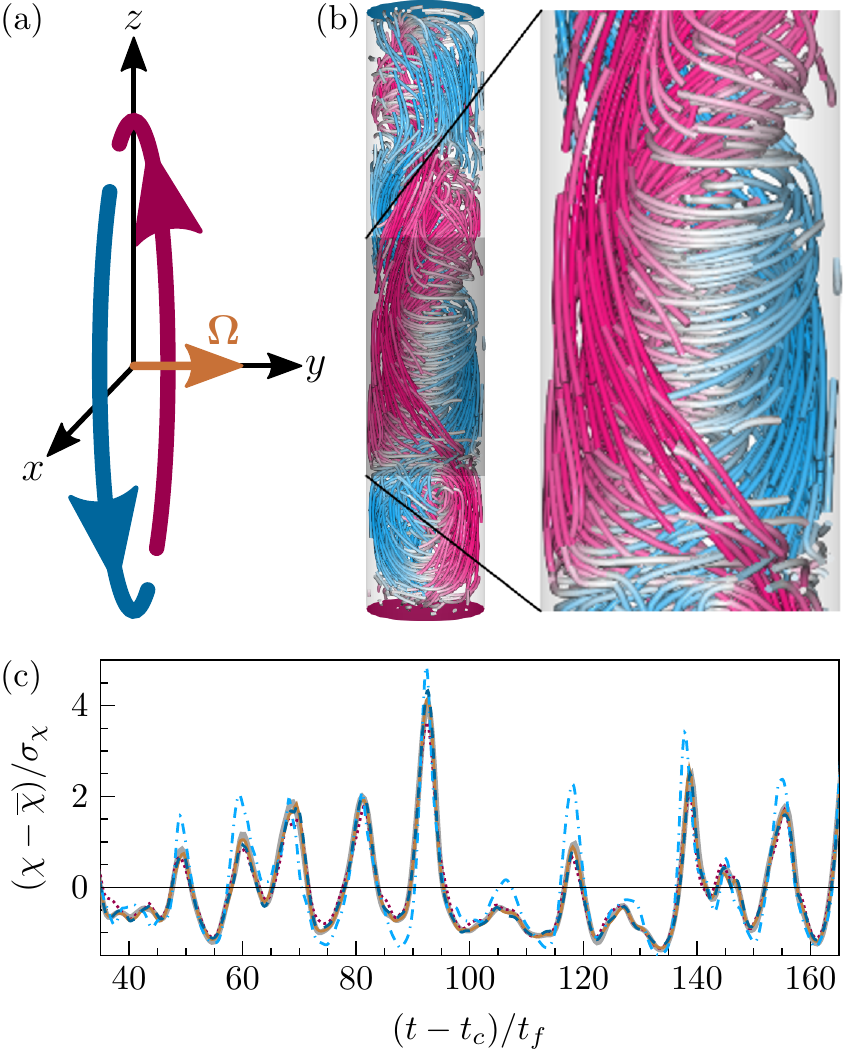}
\caption{
(a) Sketch of the primary elliptical LSC, showing the vorticity $\vec{\Omega}$ of the SRM. (b) A snapshot illustrating a strong azimuthal motion, due to the elliptical instability (colors as in Fig.~\ref{FIG1}a--d). (c) Time signal of the quantity $\chi$, which is either $\langle\Nu\rangle(t)$ (grey thick solid line), $\langle\omega^2_\text{h}\rangle(t)$ (dark blue dashed line), $\langle\omega^2_{v}\rangle(t)$ (light blue dashed-dotted line), $\langle\varepsilon_u\rangle(t)$ (red dotted line) or $\langle\varepsilon_\theta\rangle(t)$ (brown thin solid line).
Each signal is shifted by its correlation time $t_c$ with respect to $\langle\Nu\rangle(t)$.
From each quantity $\chi$ the respective mean value $\overline\chi$ is subtracted and then normalized by its standard deviation~$\sigma_\chi$.
DNS for $\Ra=5\times10^6, \Pran=0.1$, $\Gamma=1/5$.
}
\label{FIG2}
\end{figure}


\paragraph{Mechanism of the mode transitions.}
%
The elliptical instability refers to the linear instability mechanism that arises from 2D elliptical streamlines and generates a 3D flow~\cite{Kerswell2002}.
In its simplest form, the elliptical instability appears for an unbounded strained vortex in inviscid flow, $\vec{U} = (\xi-\eta) z\uvec{x}- (\xi+\eta) x\uvec{z}$, where $\uvec{x}$ and $\uvec{z}$ are the unit vectors in $x$ and $z$ directions, respectively (Fig.~\ref{FIG2}a).
The strain is denoted by $\eta$ and this vortex has a constant vorticity $\vec{\Omega} = 2\xi\uvec{y}$ and is characterized by the aspect ratio $\Gamma=\sqrt{(\xi-\eta)/(\xi+\eta)}$.
Since the SRM in a slender cylindrical cell resembles such an elliptical vortex (Fig.~\ref{FIG1}a), this instability possibly triggers its break up, and thus the emergence of the multi-roll flow modes.
Assuming that the interior of the LSC is nearly isothermal, the stability analysis of the LSC is identical to the stability analysis of an elliptical vortex \cite{Waleffe1990, Landman1987}.
The unstable mode contains vorticity along the $z$-direction.
Thus, an indicator of the elliptical instability is the growth of vorticity in the direction orthogonal to the vorticity of the elliptical flow, $\vec{\Omega}$.
In Fig.~\ref{FIG2}b a snapshot of the trajectories of passive tracer particles is shown, and a prominent azimuthal flow is visible, which twists and/or breaks up the single-roll LSC.
A necessary requirement for the elliptical instability to emerge is that the growth rate, $\sigma$, is much larger than the damping rate due to viscous dissipation, which is of the order $\nu/H^2$.
To estimate $\sigma$, it is assumed that the aspect ratio of the elliptical SRM is the same as that of the cell, hence $\Gamma=1/5$.
The vorticity, $2\xi$, of the SRM is approximated by taking the square root of the averaged horizontal enstrophy $\sqrt{\overline{\langle\omega^2_h\rangle}}\approx 7/t_f$ (TAB.~\ref{TAB:roll-properties}).
The inviscid growth rate for the aspect ratio $1/5$ is then approximated as $\sigma\approx0.3\xi$ (Fig.~1 in~\cite{Landman1987}) or $\sigma\approx1/t_f$.
However, the viscous damping is $\nu/H^2\approx1.4\times10^{-4}/t_f$ and thus about four orders of magnitude smaller than the growth rate.
Therefore, the elliptical instability is strong enough to grow.

In RBC, the following relationships of the energy dissipation rates and $\overline{\Nu}$ hold:
$\overline{\langle\varepsilon_u \rangle} = \nu\overline{\langle\omega^2\rangle}={\nu^3}H^{-4}{\Ra}{\Pran^{-2}}\left(\overline{\Nu}-1\right)$ and $\overline{\langle\varepsilon_\theta\rangle} = \kappa\Delta^2H^{-2}\overline{\Nu}$.
Although these equations are fulfilled for the time averaged quantities, their respective time series are highly correlated as well.
An example from these time series and their shift can be seen in Fig.~\ref{FIG2}c.
This temporal correlation also holds, if one considers the vertical and horizontal enstrophy components separately.
Here, we calculate the correlations in time with respect to $\Nu(t)$ to find the temporal sequence of the underlying processes.
An increase of $\Nu(t)$ is followed by an increase of the kinetic energy or $\Rey(t)$ approximately $0.55\,t_f$ later. 
After that, the thermal dissipation rate, $\varepsilon_\theta$, increases ($\approx0.76\,t_f$ later). 
Shortly after that, the horizontal enstrophy, $\omega^2_\text{h}$, increases ($\approx0.84\,t_f$ later), which is due to the strengthening of the LSC. 
Finally, the vertical enstrophy, $\omega^2_\text{v}$, increases ($\approx2\,t_f$ later), which is presumably caused by the elliptical instability.
The delay of $\approx2\,t_f$ is, compared to the mean lifetime, $\tau_n$, of a $n$-roll flow mode (Tab.~\ref{TAB:roll-properties}), of similar duration.
Note, that the kinetic energy dissipation rate, $\varepsilon_u$, which is the sum of the horizontal and vertical enstrophy, has a correlation time of $\approx1\,t_f$ which lies, as expected, in between the correlation times of each component.
The average time period of the fluctuations of $\Nu(t)$ is $T_{\Nu}\approx12\,t_f$, hence, the elliptical instability arises delayed by $\approx T_{\Nu}/6$.
During one period, the LSC can undergo several mode transitions.
This demonstrates the temporal interplay of the heat transport, circulation strength and growth of the instability.



\paragraph{Conclusions.} It was found that in laterally confined RBC flow modes with more than two rolls stacked on top of each can form.
Inside a cylindrical cell of the aspect ratio $\Gamma=1/5$ up to four rolls develop for $\Pran=0.1$ and $\Ra=5\times 10^6$.
Based on our long-term DNS, we found that the LSC, which consists of more rolls, transports heat less efficiently.
The emergence of the multi-roll flow modes is presumably caused by the elliptical instability.

\vspace{1em}
\begin{acknowledgments}
	This work is supported by the Priority Programme SPP 1881 ``Turbulent Superstructures'' of the Deutsche Forschungsgemeinschaft (DFG) under the grant Sh405/7.
	The authors acknowledge the Leibniz Supercomputing Centre (LRZ) for providing computing time.
\end{acknowledgments}

\end{document}